\documentclass[preprint,aps]{revtex4}

\usepackage{graphicx}

\begin{document}

\title{Soliton Wall Superlattice Charge-Density-Wave Phase in
Quasi-One-Dimensional Conductor (Per)$_2$Pt(mnt)$_2$}

\date{\today}

\author{Si Wu}
\author{A.G. Lebed$^*$}
\affiliation{Department of Physics, University of Arizona, 1118 E. 4th St., Tucson, AZ 85721}

\begin{abstract}

We demonstrate that the Pauli spin-splitting effects in a magnetic
field improve nesting properties of a realistic
quasi-one-dimensional electron spectrum. As a result, a high
resistance Peierls charge-density-wave (CDW) phase is stabilized in
high enough magnetic fields in (Per)$_2$Pt(mnt)$_2$ conductor. 
We show that, in low and very high magnetic fields, the Pauli
spin-splitting effects lead to a stabilization of a soliton wall
superlattice (SWS) CDW phase, which is characterized by periodically
arranged soliton and anti-soliton walls. We suggest experimental
studies of the predicted first order phase transitions between the
Peierls and SWS phases to discover a unique SWS phase. It is
important that, in the absence of a magnetic field and in a
limit of very high magnetic fields, the suggested model is equivalent
to the exactly solvable model of Brazovskii, Dzyaloshinskii, and
Kirova.
\\ \\ PACS numbers: 71.45.Lr, 74.70.Kn, 71.10.-w
\end{abstract}

\maketitle

\section{Introduction} \label{Introduction}

It is well known that the charge-density-wave (CDW) phases are generally
destroyed by a magnetic field due to the Pauli spin-splitting
effects, i.e., they are paramagnetically limited$^{1-6}$. On the
other hand, the spin-density-wave (SDW) phases are not sensitive to the Pauli
spin-splitting effects$^{2,7-12}$. In some quasi-one-dimensional
(Q1D) organic materials, the field induced dimensional crossovers in
an electron motion$^{2,7}$ can even enhance the SDW instability
and lead to a cascade of phase transitions, which is known as the
field-induced spin-density-wave (FISDW) one$^{2,7-12}$. This idea
has been also applied to the CDW phases$^{3,4}$, where the field
induced dimensional crossovers are shown to restore the CDW
instability, but only at rather low temperatures$^{4}$. Therefore,
the recent discovery of a high resistance state in Q1D materials
(Per)$_2$X(mnt)$_2$ ($X=$ Pt and Au) at high magnetic field by Graf
{\it et al}$^{13}$ is very surprising and interesting.

Originally, the above mentioned phenomenon is explained$^{13-15}$ in
terms of the dimensional crossovers effects$^{2-4,7}$. This
explanation may work for (Per)$_2$Au(mnt)$_2$, where the high
resistance state is observed only for a magnetic field, applied
perpendicular to the conducting planes, ${\bf H} \parallel {\bf c}$,
and, thus, the orbital effects play an important role. In a sister
compound (Per)$_2$Pt(mnt)$_2$, however, the high resistance state is
observed at any direction of a magnetic field$^{13,14}$. In
particular, when magnetic field is parallel to the conducting
chains, ${\bf H} \parallel {\bf b}$, the dimensional crossovers effects$^{2-4,7}$
 do not occur. Therefore, the observations of the high resistance state in
(Per)$_2$Pt(mnt)$_2$,$^{13,14}$ which is almost independent on a direction
of a magnetic field, indicates that this unique phenomenon cannot
be explained by the previous theories$^{1-4,7-10}$.

Based on the band calculation$^{16}$ and the experiments$^{17,18}$,
we have proposed a simplified but realistic Q1D model electron
spectrum to explain the main features of the phase diagram in
(Per)$_2$Pt(mnt)$_2$ conductor$^{19}$. We have demonstrated that the
Pauli spin-splitting effects improve nesting conditions for the
suggested Fermi surface and, therefore, a traditional Peierls CDW
state restores at high magnetic fields. We have also suggested a hypothesis$^{19}$ 
that, at
low and higher enough magnetic fields, a unique soliton wall
superlattice (SWS) phase may appear. This phase is characterized by two
gaps in the corresponding electron spectrum and by periodically
arranged soliton and anti-soliton walls in a real space. The
distance between these walls and values of the gaps in an energy
spectrum depend on a value of a magnetic field. 
Below, we study the phase diagram in more details. 
The main result of the present paper is a confirmation of the above
mentioned hypothesis$^{19}$.
In particular, we calculated the Landau free energy, minimize it, and
show that, indeed, the SWS phase is a ground state at low and high enough
magnetic fields.
We also demonstrate that the phase transitions between the Peierls and 
SWS phases are of the first order and suggest some experimental methods 
to discover the unique
SWS phase.

The outline of the paper is as follows. In Sec. \ref{SINe}, the spin
improved nesting phenomenon, which is crucial for understanding of
the phase diagram in (Per)$_2$Pt(mnt)$_2$ conductor, is discussed
qualitatively. Then, a transition line from metallic phase to the
CDW phase is determined in Sec. \ref{transition} by means of finite
temperature Green functions technique. In Sec. \ref{FOTran}, a free
energy relative to a metallic phase is calculated, which allows to
obtain a detailed phase diagram.

\section{Spin Improved Nesting} \label{SINe}
To begin with, let us discuss the spin improved nesting phenomenon,
which results in a stabilization of the traditional Peierls CDW
state at high enough magnetic fields in (Per)$_2$Pt(mnt)$_2$
conductor. We use the following simplified model electron spectrum, 
corresponding to four plane sheets of the Fermi surface,
\begin{equation}
\label{Spectrum}
\varepsilon_{\alpha}^{\pm}({\bf p})=\pm v_F\big[p_y\mp p_F
\pm (\Delta p/2)(-1)^{\alpha}\big].
\end{equation}
[Here $p_F$ and $v_F$ are the average Fermi momentum and the Fermi
velocity, $+(-)$ stands for right (left)
part of the Fermi surface, $\alpha=1(2)$ stands for the first
(second) conducting perylene chain, $\Delta p$ is a difference
between values of the Fermi momenta on two different conducting
chains, and $p_y$ is an electron momentum along the conducting
direction.] Note that this model is based on numerical band
calculations$^{16}$ and experimentally observed quantum interference
oscillations$^{17}$ and Landau level quantization$^{18}$. Although
the band calculations$^{16}$ indicate that the actual Fermi surface
consists of eight slightly corrugated open sheets, four of them are
almost identical to the other four, and, thus, we do not distinguish
between them. Notice that in Eq.(\ref{Spectrum}) we also ignore
electron motion in perpendicular to the chains directions. This
seems to be legitimate in (Per)$_2$Pt(mnt)$_2$ since the
corrugations of the open sheets of the Fermi surface (\ref{Spectrum})
are less than the distance between the Fermi surfaces, 
$\Delta p$.$^{16.17}$

\begin{figure}[h]
\includegraphics[width=.8\textwidth]{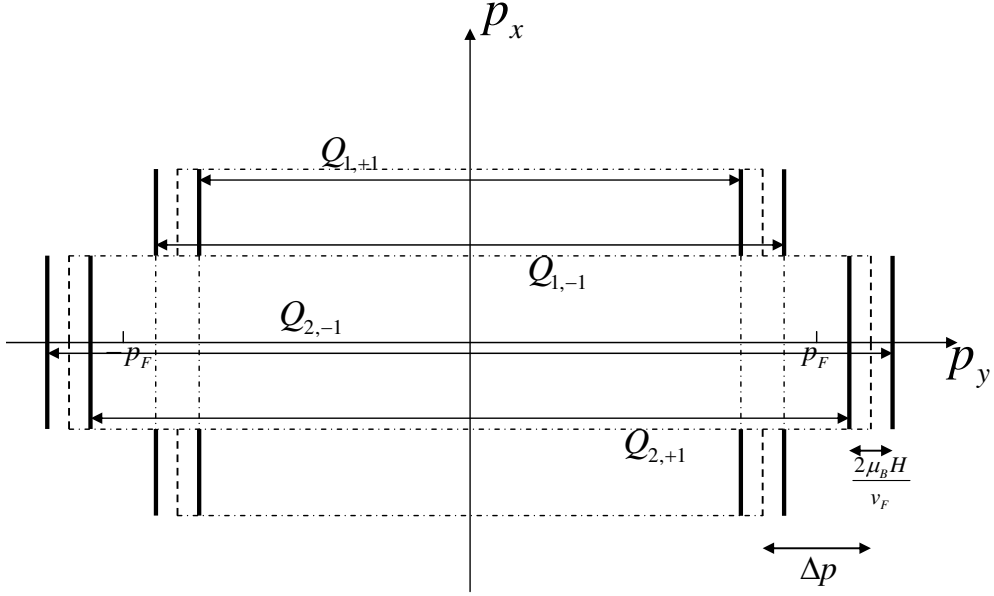}
\caption{Fermi surfaces of a Q1D conductor with two conducting
chains in a magnetic field. The original four sheets are split into
eight ones [see Eq. (\ref{Split})] and, thus, a competition between the CDW
phases is characterized by four different nesting vectors,
$Q_{1,+1}$,$Q_{1,-1}$,$Q_{2,+1}$,$Q_{2,-1}$ [see Eqs.
(\ref{Nesting2})]. At magnetic field, $H_p^*=\Delta pv_F/2\mu_B$,
two nesting vectors coincide, $Q_{1,-1}=Q_{2,+1}=2p_F$, with a half
of the original Fermi surface being nested. This results in a
restoration of the Peierls CDW phase at high magnetic fields [see
Eqs.(\ref{CriticalField}),(\ref{Transition1}) and Fig.\ref{PD}].}
\label{fig01}
\end{figure}

In a magnetic field, the electron spectrum (\ref{Spectrum}) is
split into eight sheets,
\begin{equation}
\label{Split}
\varepsilon_{\alpha\sigma}^{\pm}({\bf p})=\pm
v_F\big[p_y\mp p_F \pm (\Delta p/2)(-1)^{\alpha}\big]-\sigma\mu_BH \
,
\end{equation}
where $\sigma=\pm 1$ is a spin component of the electron along a
direction of a magnetic field and $\mu_B$ is the Bohr magneton. As
shown in Fig.\ref{fig01}, there exist four different nesting vectors
competing with each other, $Q_{1,+1}$, $Q_{1,-1}$, $Q_{2,+1}$, and
$Q_{2,-1}$. Note that the Peierls CDW instability, which results
from pairing of electrons and holes with the same spins and with
momenta difference $2p_F$, is, thus, paramagnetically limited. This is
clearly seen from Fig.\ref{fig01}. Indeed, two original nesting
vectors,
\begin{equation}
\label{Nesting1} Q=2p_F\pm\Delta p,
\end{equation}
are split in the presence of a magnetic field into four ones,
\begin{equation}
\label{Nesting2} Q_{\alpha\sigma}=2p_F+q_{\alpha\sigma},\ \ \ \
q_{\alpha\sigma}= (-1)^{\alpha} \Delta p  -2\sigma\mu_BH/v_F,
\end{equation}
which decreases instability to a formation of the CDW. Moreover, the
existence of these four nesting vectors may even correspond to the
appearance of several energy gaps in an electron spectrum at high
values of the parameters $\Delta p$ and $2\mu_BH/v_F$. Our
theoretical results, as shown below, confirm the appearance of the
SWS phase with two energy gaps, which is in a qualitative agreement
with a general theory of solitons and soliton
superstructures$^{20-24}$.

However, according to Fig.\ref{fig01} and Eq. (\ref{Nesting2}), at
some critical value of a magnetic field,
\begin{equation}
\label{CriticalField} H^*_p=\Delta pv_F/2\mu_B \ ,
\end{equation}
two of the nesting vectors coincide, $Q_{1,-1}=Q_{2,+1}=2p_F$, and
the number of nesting vectors decreases to three, which improves
the nesting conditions. Moreover, according to Eq. (\ref{Nesting2}), a
half of the original sheets of Fermi surface are nested with
$Q=2p_F$. Therefore, we expect a restoration of the traditional
Peierls CDW phase with one gap in an electron spectrum in the
vicinity of $H \approx H_p^*$ (see Figs. \ref{PD}, \ref{Gap}). 
These statements are confirmed later by our theoretical 
analysis.

\section{Metal-CDW Transition Line} \label{transition}

Now let us consider the CDW phase with nesting vector,
\begin{equation}
\label{NestingVector}
{\bf Q} = (0,2p_F+q,0),
\end{equation}
and the CDW order parameter,
\begin{equation}
\label{OrderParameter}
\Delta_{CDW}(x) = \Delta_q{\rm e}^{i(2p_F+q)x} +
\Delta_q^*{\rm e}^{-i(2p_F+q)x}.
\end{equation}
Below, we use the finite temperature Green function method$^{25}$ to
study the metal-CDW phase transition line. We consider the following
standard mean field Hamiltonian,
\begin{eqnarray}
\label{Hamiltonian}
\hat H &=&\sum_{\alpha=1,2}\sum_{\sigma= \pm 1}\sum_{\xi}\Big\{
a_{\alpha\sigma}^{\dagger}(\xi)a_{\alpha\sigma}(\xi)
\big[\varepsilon_{\alpha\sigma}^+(\xi)-\mu\big]+
b_{\alpha\sigma}^{\dagger}(\xi)b_{\alpha\sigma}(\xi)
\big[\varepsilon_{\alpha\sigma}^-(\xi)-\mu\big]\Big\}\nonumber\\
&+&\sum_{\alpha=1,2}\sum_{\sigma= \pm 1}\sum_{\xi}\Big\{
a_{\alpha\sigma}^{\dagger}(\xi)b_{\alpha\sigma}(\xi-q) \Delta_q+
b_{\alpha\sigma}^{\dagger}(\xi)a_{\alpha\sigma}(\xi+q)\Delta_q^*\Big\},
\end{eqnarray}
where
\begin{equation}
\label{FieldOperator}
\Psi_{\alpha \sigma} (x)= \exp(-i p_Fx) \sum_{\xi} e^{i \xi x}
b_{\alpha \sigma}(\xi)+ \exp(i p_Fx) \sum_{\xi} e^{i \xi x} a_{\alpha \sigma} (\xi)
\end{equation}
is a field operator of an electron, with $a_{\alpha\sigma}(\xi)$ and
$b_{\alpha\sigma}(\xi)$ being electron annihilation operators near
the right and left sheets of the Fermi surface, correspondingly.

Following the same approach as in the theory of superconductivity,
we define the normal and anomalous (Gor'kov) Green functions,
\begin{equation}
\label{GreenFunction}
G_{\alpha\sigma}^{++}(\xi, \tau)=-\langle
T_{\tau}a_{\alpha\sigma}(\xi,\tau)
a_{\alpha\sigma}^{\dagger}(\xi,0)\rangle,\nonumber\\
G_{\alpha\sigma}^{-+}(\xi, \tau)=-\langle
T_{\tau}b_{\alpha\sigma}(\xi-q,\tau)
a_{\alpha\sigma}^{\dagger}(\xi,0)\rangle,
\end{equation}
and derive the corresponding equations of motion,
\begin{equation}
\label{Motion}
\Big(i\omega_n-\big[\varepsilon_{\alpha\sigma}^+(\xi)-\mu\big]\Big)
G_{\alpha\sigma}^{++}(\xi,i\omega_n)-\Delta_q
G_{\alpha\sigma}^{-+}(\xi,i\omega_n)=1,
\end{equation}
\begin{equation}
\label{2Motion}
\Big(i\omega_n-\big[\varepsilon_{\alpha\sigma}^-(\xi-q)-\mu\big]\Big)
G_{\alpha\sigma}^{-+}(\xi,i\omega_n)-\Delta^*_q
G_{\alpha\sigma}^{++}(\xi,i\omega_n)=0.
\end{equation}
In this case, the gap is self-consistently determined by
\begin{equation}
\label{GapEquation}
\Delta^*_q=-g^2\sum_{\alpha=1,2}\sum_{\sigma= \pm 1}\sum_{\xi}T\sum_{\omega_n}
G_{\alpha\sigma}^{-+}(\xi,i\omega_n),
\end{equation}
where $\omega_n=2\pi T(n+1/2)$ is the Matsubara frequency$^{25}$.

Solution of a linearized variant of Eqs.
(\ref{Motion})-(\ref{GapEquation}) gives us the following transition
line between the metallic and CDW phases,
\begin{equation}
\label{Transition1}
\ln
\biggl( \frac{T_{c0}}{T_c} \biggl)=\frac{1}{4}\sum_{\alpha=1,2}
\sum_{\sigma=\pm 1}
\sum_{n=0}^{\infty}\frac{v_F^2(q-q_{\alpha\sigma})^2/(4\pi T_c)^2}
{(n+\frac{1}{2})\big[(n+\frac{1}{2})^2+v_F^2(q-q_{\alpha\sigma})^2/(4\pi
T_c)^2\big]},
\end{equation}
where $q_{\alpha\sigma}$ are given by Eq. (\ref{Nesting2}). Note
that Eq. (\ref{Transition1}) can be rewritten in a more terse way
using the so-caled $\psi$ function$^{26}$,
\begin{equation}
\label{Transition2}
\ln
\biggl( \frac{T_{c0}}{T_c} \biggl)=
\frac{1}{4}
\sum_{\alpha=1,2}
\sum_{\sigma=\pm 1}
 \biggl( \frac{1}{2}
 \psi \biggl[  \frac{1}{2} + i \frac{v_F (q-q_{\alpha\sigma})}{4 \pi T_c}
\biggl]
+  \frac{1}{2} \psi \biggl[  \frac{1}{2} - i \frac{v_F (q-q_{\alpha\sigma})}{4 \pi T_c}
\biggl]
- \psi \biggl[ \frac{1}{2} \biggl]
\biggl) \ .
\end{equation}
Each of these equations defines the metal-CDW transition line. In
particular, they determine a transition temperature $T_c$ for
electron spectrum (\ref{Spectrum}) in the presence of a magnetic
field using a transition temperature $T_{c0}$ value, corresponding
to ideal nesting conditions (i.e., $H=0$ and $\Delta p = 0$). Note
that a competition between the four nesting vectors of Eq.
(\ref{Nesting2}), discussed in Sec. \ref{SINe}, is directly seen in
Eqs. (\ref{Transition1}), (\ref{Transition2}).

\begin{figure}[h]
\includegraphics[width=.8\textwidth]{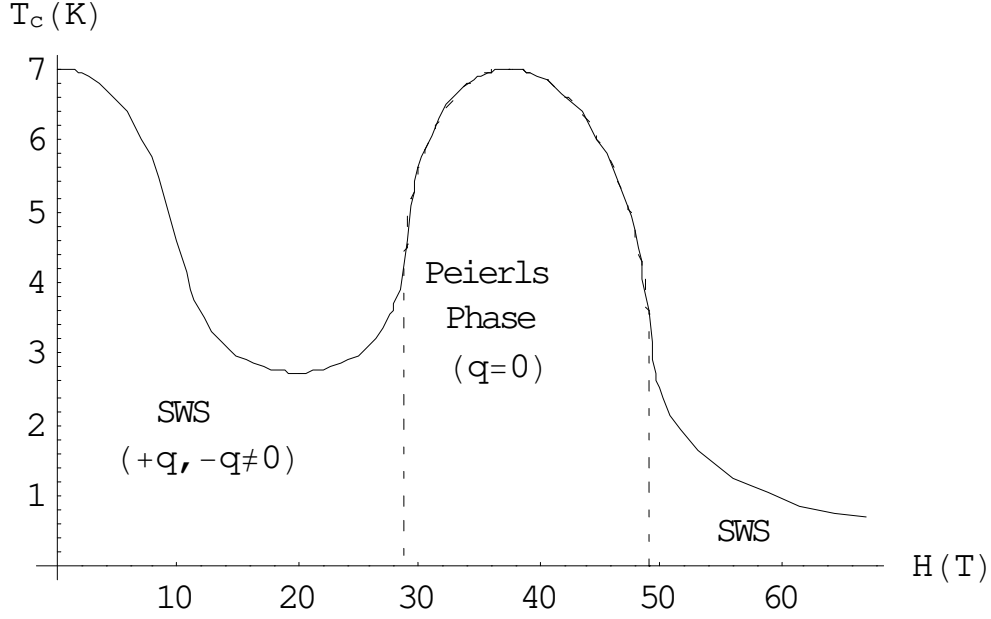}
\caption{Hypothetical phase diagram$^{19}$ of
(Per)$_2$Pt(mnt)$_2$ Q1D conductor in a magnetic field. Solid line
is a phase transition line between the metallic and CDW phases,
calculated from Eq. (\ref{Transition1}). The dotted lines separate
the Peierls and SWS phases. For confirmation of this phase diagram,
see Sec. IV and Fig.5.} \label{PD}
\end{figure}

\begin{figure}[h]
\includegraphics[width=.8\textwidth]{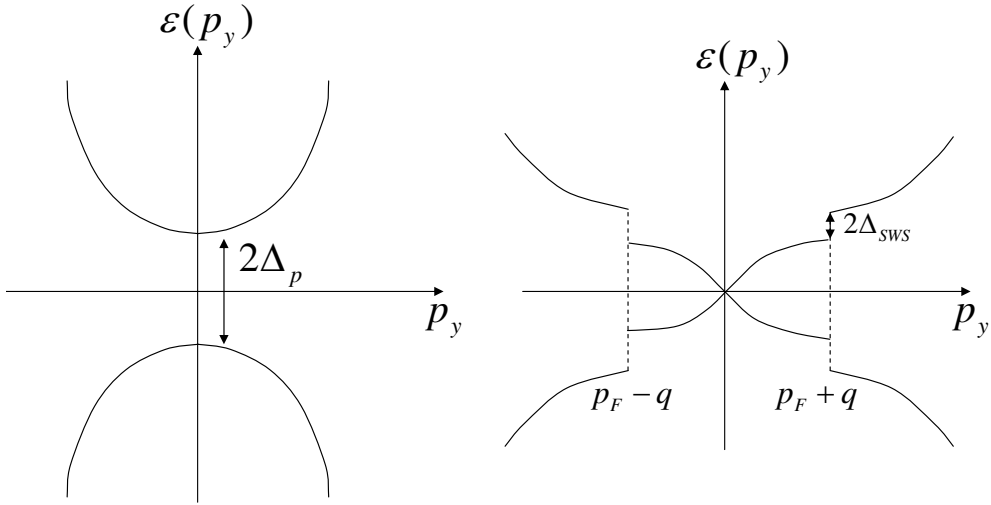}
\caption{Electron spectrum of the Peierls phase (left) has one
energy gap, $\Delta_p$, whereas the SWS phase (right) is
characterized by two smaller energy gaps, $\Delta_{SWS}$. This
results in different optical and thermodynamical properties of the
Peierls and SWS phases.} \label{Gap}
\end{figure}

Numerical solutions of Eq. (\ref{Transition1}) is presented in Fig.
\ref{PD}, where we use a value of the parameter $\Delta p v_F=60$K,
determined from a theoretical analysis of the experimentally
observed quantum magnetic oscillations$^{17}$. As seen from Fig.
\ref{PD}, the Peierls phase is stabilized at high enough magnetic
fields, $29{\rm T}<H<49{\rm T}$. 
At very high magnetic fields, $H>49$T, and low magnetic fields, $H<29$T, an incommensurate CDW  phase is
shown to be a ground state. 
We suggest a hypothesis$^{19}$ that this incommensurate phase actually
corresponds to the SWS ground state.
The latter statement is proved by an
analysis of the Landau free energy in Sec. \ref{FOTran} (see also
Fig. \ref{ModifiedPD}). We point out that the calculated in this section
metal-CDW phase transition line is in very good qualitative and quantitative 
agreements with the observed one$^{13,14}$.

\section{First Order Phase Transitions}
\label{FOTran}

It is known that, close to the metal-CDW second order phase
transition line, the order parameter is vanishingly small and the
Landau theory of the second order phase transitions can be applied.
Note that the SWS phase, in the vicinity of the metal-CDW phase
transition line, is characterized by the following order
parameter,$^{21-24}$
\begin{equation}
\Delta_{SWS}(x)=\Delta\cos(qx)\cos(2p_Fx),
\end{equation}
which corresponds to mixing of two order parameters (7) with $+q$, $\Delta_q$,
and $-q$, $\Delta_{-q}$,
where $q\neq 0$. Therefore, below, we derive the Landau free energy up to the
fourth order terms in $\Delta_q$ and $\Delta_{-q}$ and study the mixing of these
order parameters in the SWS phase.

\subsection{Free Energy Correction}

In this sub-section, we consider the following improved Hamiltonian,
\begin{equation}
\label{FullHamiltonian}
\hat H = \hat H_0 + \hat H_I,
\end{equation}
where $\hat H_0$ is a kinetic energy of free electrons and,
\begin{eqnarray}
\label{HI}
\hat H_I & = & \sum_{\zeta}\sum_{\alpha\sigma}
\Big\{\Delta_q a^{\dagger}_{\alpha\sigma}(\xi+q)b_{\alpha\sigma}(\xi)
+\Delta_q^* b^{\dagger}_{\alpha\sigma}(\xi)a_{\alpha\sigma}(\xi+q) \nonumber\\
&&+\Delta_{-q}a^{\dagger}_{\alpha\sigma}(\xi-q)b_{\alpha\sigma}(\xi)
+\Delta_{-q}^*b^{\dagger}_{\alpha\sigma}(\xi)a_{\alpha\sigma}(\xi-q)\Big\},
\end{eqnarray}
is a mean-field Hamiltonian for interactions between the electrons
and the CDW lattice deformation. In contrast to the Hamiltonian of Sec.
\ref{transition}, a possible mixing of the order parameters,
$\Delta_q$ and $\Delta_{-q}$, has been taken into account in Eq.
(\ref{HI}).

Below, we apply to Hamiltonian (18) a diagram technique for a thermodynamic 
potential, described, for example, in Ref. [25]. This allows us to determine
the Landau free energy up to the fourth order terms in the order
parameters, $\Delta_q$ and $\Delta_{-q}$,
\begin{equation}
\label{FourthOrder}
\Delta F = \gamma(|\Delta_q|^2+|\Delta_{-q}|^2)
+ \eta _1(|\Delta_q|^4+|\Delta_{-q}|^4) + \eta_2|\Delta_q\Delta_{-q}|^2,
\end{equation}
where details of our calculation of the coefficients, $\gamma$,
$\eta_1$, and $\eta_2$, can be found in Appendix A. [Note that in
the paper we consider a so-called incommensurate model of
(Per)$_2$Pt(mnt)$_2$ electron spectrum. It is an appropriate
approximation for our calculations since very weak 1/4
commensurability effects are easily destroyed by small but non-zero
corrugations of the Q1D Fermi surfaces in (Per)$_2$Pt(mnt)$_2$.]

As known, the coefficient $\gamma$ determines the metal-CDW
transition line if setting to zero. It is positive in the metallic
and negative in CDW phases. Therefore, to determine the CDW ground
state, we need to minimize the free energy (\ref{FourthOrder}) for
$\gamma<0$. In non-SWS phase, where only one order parameter is
non-zero (e.g., $\Delta_q\neq 0$ and $\Delta_{-q}=0$), the
minimization procedure results in
\begin{equation}
\label{NonSoliton} \Delta F_{NS} = -\frac{\gamma^2}{4\eta_1}.
\end{equation}
In the SWS phase, where there is a mixing of the order
parameters, $\Delta_q$ and $\Delta_{-q}$, the free energy is
\begin{equation}
\label{Soliton} \Delta F_{S} = -\frac{\gamma^2}{2\eta_1+\eta_2}.
\end{equation}
Comparing Eqs. (\ref{NonSoliton}), ({\ref{Soliton}}), we find the
following condition for the appearance of the SWS phase:
\begin{equation}
2\eta_1 > \eta_2.
\end{equation}
Below, we use Eqs. (\ref{NonSoliton}), (\ref{Soliton}) for the
Landau free energy to study the CDW phase in more detail.

\subsection{Results}
\label{Results}

In this sub-section, the phase transition lines from Peierls phase to
the SWS phase are numerically calculated by means of Eqs.
(\ref{NonSoliton}), (\ref{Soliton}). We discuss in detail the phase
transition lines in the vicinity of $H\approx 49$T. In the vicinity
of $H\approx 29$T, the phase diagram is qualitatively similar to
that in the vicinity of $H\approx 49$T.

\begin{figure}[h]
\includegraphics[width=.8\textwidth]{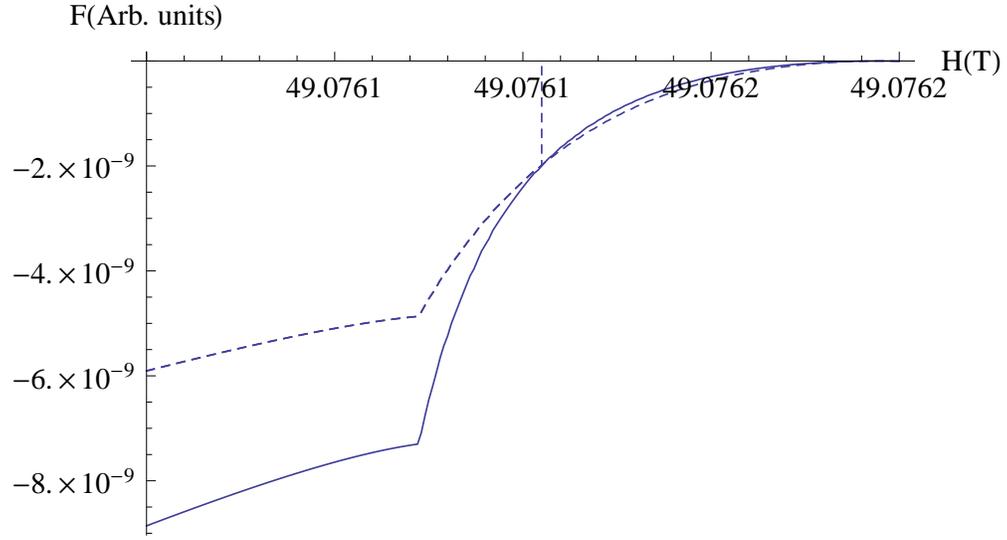}
\caption{The Landau free energies (\ref{NonSoliton}),
(\ref{Soliton}) in the vicinity of a metal-CDW transition line are
calculated in both the SWS and non-SWS phases. Solid and dashed
lines stand for the free energies of the non-SWS and SWS phases,
respectively. } \label{FreeEnergy}
\end{figure}

\begin{figure}[h]
\includegraphics[width=.8\textwidth]{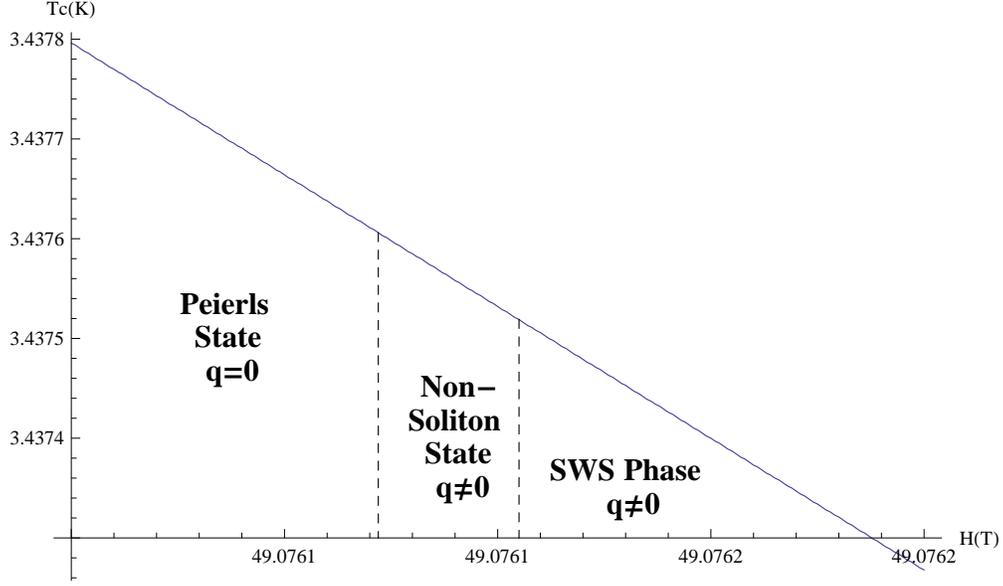}
\caption{The detailed phase diagram in the vicinity of $H\approx
49$T. Solid line: a second order phase transition line between the metallic and
CDW phases. Dashed lines: left, the first-order phase transition
line from the Peierls state to the incommensurate non-SWS phase with
non-trivial nesting vector; right, the first-order phase transition
line from the incommensurate non-SWS phase to the SWS phase.
Due to extremely narrow region of a stability of the incommensurate non-SDW
phase, we expect that there exist a direct first order phase transition from the
Peierls to SWS phases.}
\label{ModifiedPD}
\end{figure}

The results of the numerical evaluations of the Landau free energies are shown in
Fig. \ref{FreeEnergy}, where corrections for both the SWS and
non-SWS phases are calculated for the same temperature which is
slightly below the metal-CDW transition line. This guarantees a
validity of the Landau expansion for the free energies. As seen, at
$H<49.076106$T, the non-SWS phase has lower free energy, but for
higher magnetic fields, the SWS phase is a ground state. Therefore, a
true free energy curve has a discontinuity in its slope at the point
of a transition from non-SWS to SWS phases, which corresponds to
the first order phase transition (see right dashed line in Fig.
\ref{ModifiedPD}). We also note that at $H=49.076072$T, there is a
kink in the free energy line, which corresponds to another
first-order phase transition. Our detailed numerical calculations
show that, between these two first order transitions, there exists a
new non-SWS state, which is characterized by an incommensurate
nesting vector, $Q\neq 2p_F$.

A detailed phase diagram in the vicinity of $H\approx49$T is shown
in Fig. \ref{ModifiedPD}. Starting from the Peierls CDW phase, as
magnetic field increases, the ground state first becomes a
non-SWS CDW state with nontrivial nesting vector, then the system
enters into the SWS CDW phase. However, we point out that
numerically the region of a stability of the non-SWS
incommensurate phase is extremely narrow. Therefore, we expect that
thermodynamical fluctuations and hysteresis will result in a direct
first order phase transition from the Peierls into SWS phases 
(see Fig. \ref{PD}).

\section{Conclusion}
To summarize, we have suggested an explanation of the experimentally
observed high resistance state in Q1D organic conductor (Per)$_2$Pt(mnt)$_2$. 
The calculated phase diagram is in good qualitative and quantitative agreements 
with the existing experiments$^{13,14}$. We have also predicted the existence of a unique SWS phase, which is characterized by two energy gaps in its electron spectrum and corresponds to periodically arranged soliton and anti-soliton walls. Our detailed calculations of the Landau free energy demonstrates that there is an incommensurate CDW phase between the commensurate Peielrs and SWS phases. Nevertheless, an area of the stability of the incommensurate phase is shown to be numerically extremely narrow, therefore, we suggest that there is a direct first order phase transition from the commensurate Peierls into SWS phases. 
It is important that the calculated in the paper magnetic field dependence of the Landau free energy of different CDW phases is due to a pure spin contribution. Therefore, the above mentioned first order phase transition can be detected as a divergence of the Knight shift value at the corresponding magnetic field. The SWS phase can be also discovered by a detection of two energy gaps by some infra-red measurements. Indirect confirmation of the existence of the SWS phase in (Per)$_2$Pt(mnt)$_2$ is already provided by the measurements of an activation gap$^{13}$, where it is shown that the activation gap becomes very small at low and very high magnetic fields. This fact is in an agreement with the electron energy spectrum of the SWS phase (see Fig.3), which is characterized by two relatively small energy gaps$^{21-24}$, although more detailed experimental analysis is needed to make a firm statement. And finally, we suggest neutron and x-rays diffraction experiments in (Per)$_2$Pt(mnt)$_2$ to detect the predicted periodic superstructure of soliton and anti-soliton walls directly, which is the main characteristics of the predicted SWS CDW phase.

\section{Acknowledgement}

One of us (A.G.L.) is thankful to N.N. Bagmet, J.S. Brooks, P.M. Chaikin, N. Harrison, M.V. Kartsovnik, and J.S. Singleton for useful discussions. This work was supported by the NSF under Grant  DMR-0705986.

\appendix*
\section{Calculation of $\gamma$ and $\eta 's$}

In this appendix, we give an outline of the calculations of the
coefficients $\gamma$, $\eta_1$ and $\eta_2$ appearing in the Landau
free energy expansion near the metal-CDW phase transition line (see
Eq. (\ref{FourthOrder})).

Feynman diagrams contributing to the second order corrections
to the free energy$^{25}$ with respect to $\Delta_q$ and $\Delta_{-q}$ are
shown Fig. 6. As a result, we obtain:
\begin{equation}
\label{Gamma}
\gamma =  2\pi \sum_{\alpha\sigma}
\sum_{\omega_n}
\int\frac{{\rm d}\xi}{2\pi}G^{++}_{\alpha\sigma}(i\omega_n, \xi +q, H)
G^{--}_{\alpha\sigma}(i\omega_n, \xi, H).
\end{equation}
Mathematical technique for evaluation of Eq. (\ref{Gamma}) is
standard$^{25}$, which results in
\begin{equation}
\label{GammaResult1}
\gamma = -\frac{1}{2v_FT}\sum_{\alpha\sigma}\sum_{n>0}\Big\{
\frac{1}{\displaystyle n+\frac{1}{2}+\frac{iv_F(q-q_{\alpha\sigma})}{4\pi T}}
+c.c.\Big\} .
\end{equation}
In Eq. (\ref{GammaResult1}), wave vectors $q_{\alpha\sigma}$ are
defined by Eq. (\ref{Nesting2}); c.c. stands for a complex conjugate
value. After simple calculations, it can be shown that Eq.
(\ref{GammaResult1}) is equivalent to:
\begin{equation}
\label{GammaResult2}
\gamma = \ln
\biggl( \frac{T_{c0}}{T} \biggl)-\frac{1}{4}\sum_{\alpha\sigma}
\sum_{n=0}^{\infty}\frac{v_F^2(q-q_{\alpha\sigma})^2/(4\pi T)^2}
{(n+\frac{1}{2})\big[(n+\frac{1}{2})^2+v_F^2(q-q_{\alpha\sigma})^2/(4\pi
T)^2\big]}.
\end{equation}
By setting $\gamma$ to zero, we obtain the metal-CDW second order 
phase transition line, Eq. (\ref{Transition1}).

The fourth order terms can be also evaluated by using perturbation
theory for a thermodynamic potential$^{25}$. As it can be shown,
$|\Delta_q|^4$ and $|\Delta_{-q}|^4$ terms, corresponding to two
diagrams in Fig. 7, have the same coefficients and each such diagram 
is characterized by a weighting factor 2, therefore,
\begin{equation}
\label{Eta1}
\eta_1 = \frac{1}{32v_F\pi^2T^5}
\sum_{\alpha\sigma}\sum_{n>0}\Big\{
\frac{1}{\bigg[\displaystyle n+\frac{1}{2}+\frac{iv_F(q-q_{\alpha\sigma})}
{4\pi T}\bigg]^3}+c.c.\Big\}.
\end{equation}
Two diagrams, corresponding to $|\Delta_q\Delta_{-q}|^2$ term (see Fig. 8),
are equal and each of them has a weighting factor 4, therefore,
\begin{eqnarray}
\label{Eta2}
\eta_2 &=& - \frac{i}{8qv_F^2\pi T}
\sum_{\alpha\sigma}\sum_{n>0}\Big\{
\frac{1}{\bigg[\displaystyle n+\frac{1}{2}+\frac{iv_F(q-q_{\alpha\sigma})}
{4\pi T}\bigg]^2}-\frac{1}{\bigg[\displaystyle n+\frac{1}{2}-
\frac{iv_F(q-q_{\alpha\sigma})}
{4\pi T}\bigg]^2}\nonumber\\
&&-\frac{1}{\bigg[\displaystyle n+\frac{1}{2}-
\frac{iv_F(q+q_{\alpha\sigma})}
{4\pi T}\bigg]^2}+\frac{1}{\bigg[\displaystyle n+\frac{1}{2}+
\frac{iv_F(q+q_{\alpha\sigma})}
{4\pi T}\bigg]^2}\Big\}.
\end{eqnarray}

$^*$Also Landau Institute for Theoretical Physics,
2 Kosygina Street, Moscow, Russia.

\begin{figure}[h]
\includegraphics[width=.8\textwidth]{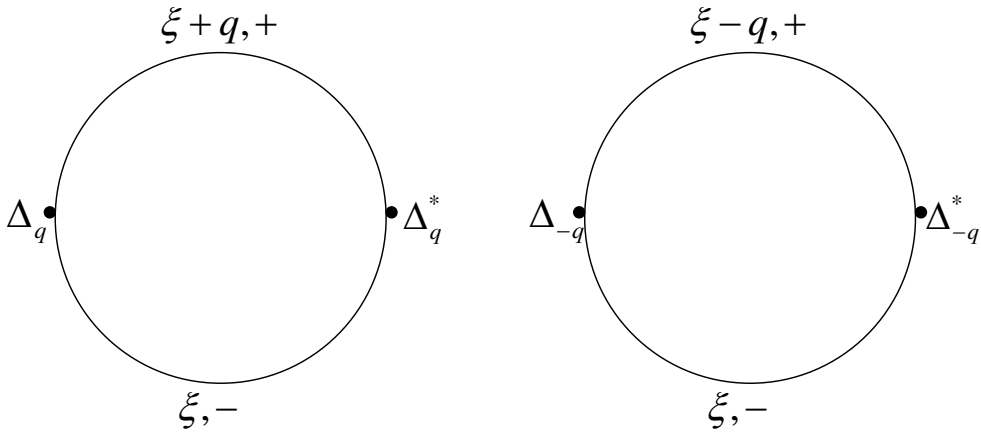}
\label{Second} \caption{Second order diagrams, corresponding to
the terms, $|\Delta_q|^2$ and $|\Delta_{-q}|^2$, in the Landau free
energy (19).}
\end{figure}

\begin{figure}[h]
\includegraphics[width=.8\textwidth]{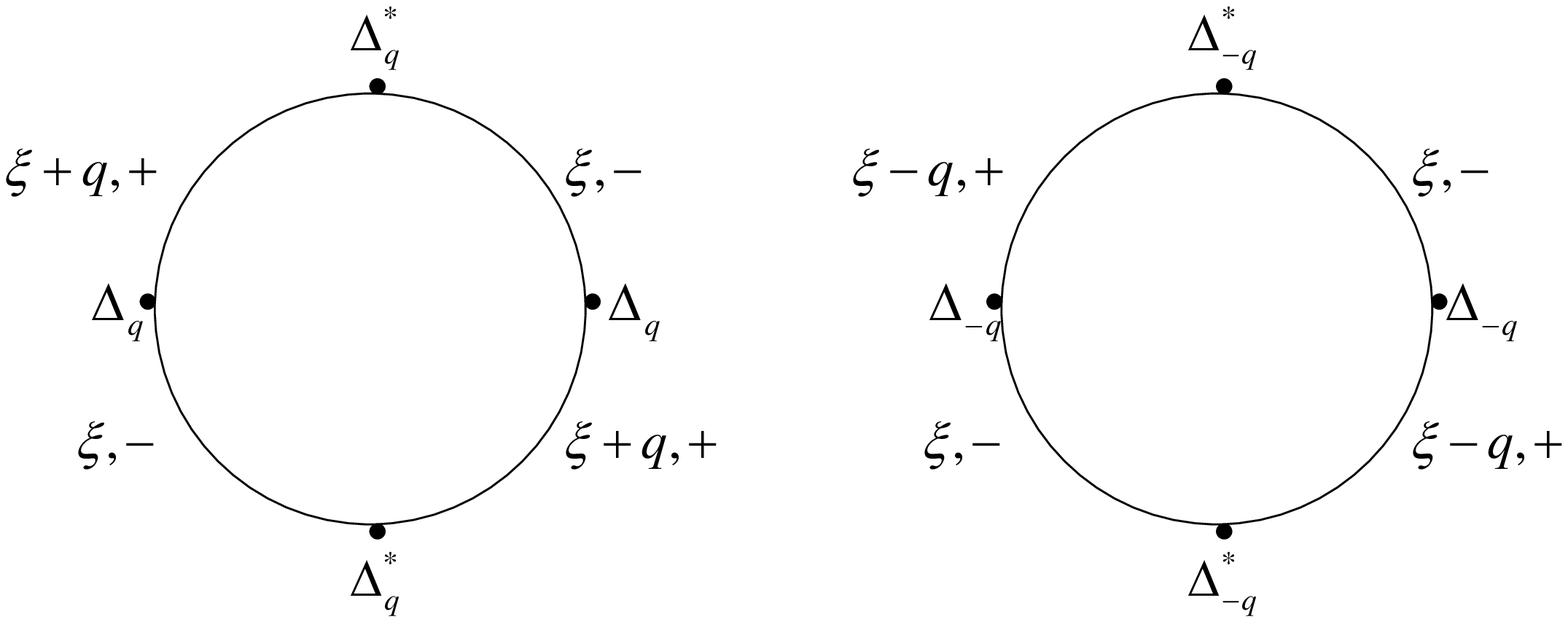}
\label{Fourth1} \caption{Fourth order diagrams, corresponding to
the terms, $|\Delta_q|^4$ and $|\Delta_{-q}|^4$, in the Landau free
energy (19).}
\end{figure}

\begin{figure}[h]
\includegraphics[width=.8\textwidth]{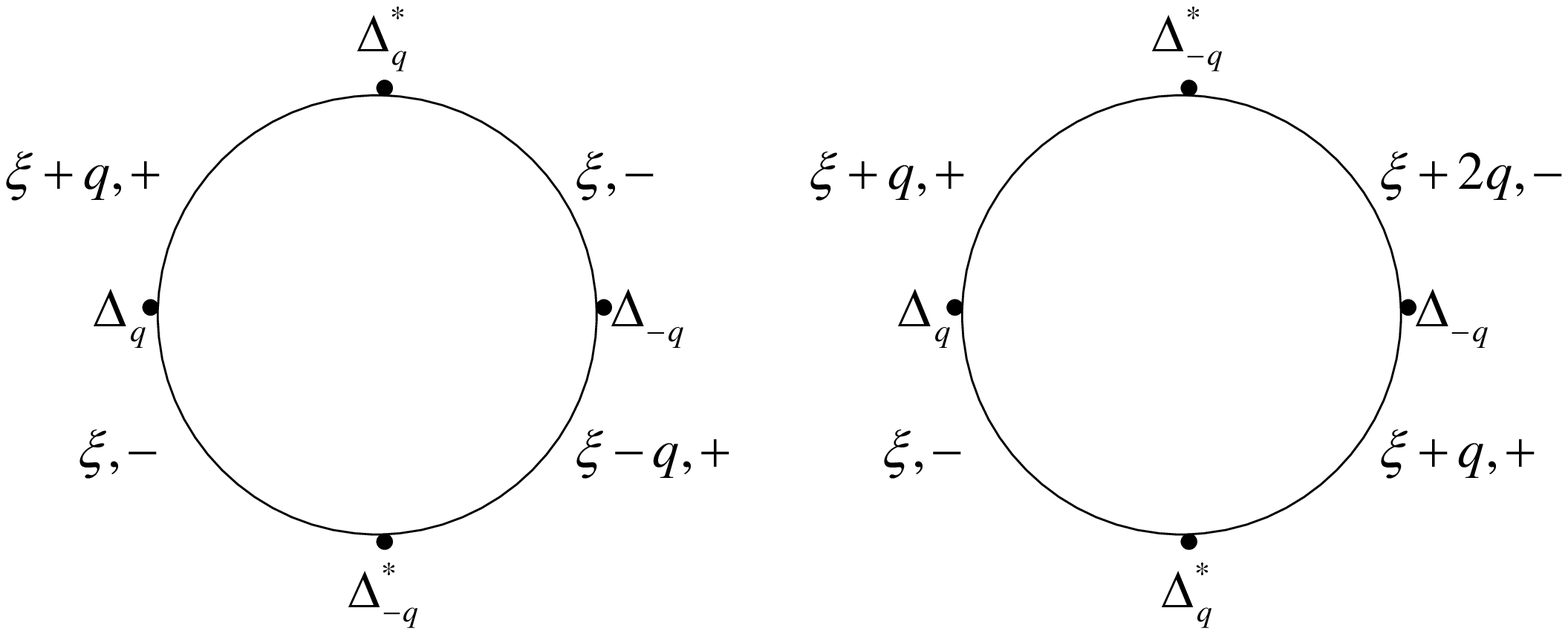}
\label{Fourth2} \caption{Fourth order diagrams, corresponding to
the term, $|\Delta_q\Delta_{-q}|^2$, in the Landau free
energy (19).}
\end{figure}


\begin{thebibliography}{99}
\bibitem{comment3}
We stress that our model is equivalent to that in Ref.\cite{Brazovskii}
in two limiting cases: at zero magnetic field, $\mu_BH/\Delta pv_F
\rightarrow 0$, and at high magnetic fields, $\mu_BH/\Delta pv_F
\rightarrow \infty$. Therefore, at high enough and low enough magnetic
fields, the SWS phase is characterized by two energy gaps at any
temperature in accordance with Ref.\cite{Brazovskii}. As shown in the
paper, the SWS phase is characterized by two energy gaps in the vicinity
of the metal-SWS phase transition line for the values of the parameters,
$T_c(H=0)=7$K and $\Delta pv_F=60$K. Note that, for higher value of
the parameter $\Delta pv_F$, the SWS phase may be characterized by four
gaps in its electron spectrum.
\end{thebibliography}
\end{document}